\documentclass{aastex631}

\newcommand{\zmax}{\langle z_{\mathrm{max}} \rangle}
\newcommand{\Rmax}{\langle R_{\mathrm{max}} \rangle}

\def\gtorder{\mathrel{\raise.3ex\hbox{$>$}\mkern-14mu
    \lower0.6ex\hbox{$\sim$}}}
\def\ltorder{\mathrel{\raise.3ex\hbox{$<$}\mkern-14mu
    \lower0.6ex\hbox{$\sim$}}}

\newcommand{\ZZ}[1]{}  % to sort bibs with same authors, same year correctly.

\usepackage{graphicx}	% Including figure files
\usepackage{cancel}     % for striking words
\usepackage{amsmath}	% Advanced maths commands
\usepackage{soul}
\usepackage{multirow}
\usepackage{hhline}

\begin{document}

\title[Resonant Cooling in Barred Galaxies]{Metal-Poor Stars in the MW Disk:\\ Resonant Cooling of Vertical Oscillations of Halo Stars in Barred Galaxies}

\author[0000-0001-9940-1522]{Xingchen Li}
\affiliation{Department of Physics \& Astronomy, University of Kentucky, Lexington, KY 40506, USA}

\author[0000-0002-1233-445X]{Isaac Shlosman}
\affiliation{Department of Physics \& Astronomy, University of Kentucky, Lexington, KY 40506, USA}
\affiliation{Theoretical Astrophysics, School of Sciences, Osaka University, Osaka 560-0043, Japan}

\author[0000-0002-0980-3622]{Daniel Pfenniger}
\affiliation{University of Geneva, Geneva Observatory, ch. Pegasi 51, 1290 Versoix, Switzerland}

\author[0000-0002-1005-6441]{Clayton Heller}
\affiliation{Department of Biochemistry, Chemistry \& Physics, Georgia Southern University, Statesboro, GA 30460, USA}

\correspondingauthor{Xingchen Li, Isaac Shlosman}
\email{xingchen.li@uky.edu, isaac.shlosman@uky.edu}



\begin{abstract}
Using numerical simulations of barred disk galaxy embedded in nonspinning and spinning dark matter (DM)  halos, we present a novel mechanism of `cooling' the vertical oscillations of DM particles, which acquire the disk kinematics. The underlying mechanism consists of resonant interactions between halo particles and the stellar bar, facilitated by chaotic phase space of the system. The cooling mechanism acts both on dynamical and secular timescales, from $\sim 0.5$\,Gyr to few Gyr. The stellar bar acts to absorb kinetic energy of the vertical motions. Using Milky Way-type stellar halo, we estimate the population of metal-poor disk stars trapped by the MW disk and analyze its kinematics. We find that population of metal-poor MW disk stars with $|z|\ltorder 3$\,kpc detected  by the Gaia DR3 and other surveys can have their origin in the stellar halo. The cooled population also migrates radially outwards by exchanging energy and angular momentum with the spinning bar, and prograde-moving stars have a different distribution from the retrograde ones. Next, we have calculated the ratio of the prograde-to-retrograde orbits of the cooled population and found that this ratio varies radially, with the fast-spinning stellar halo resulting in the shallower radial increase of this ratio outside of the corotation. The nonspinning stellar halo shows a monotonic increase of this ratio with radius outside the corotation. Together with analyzed radial migration of these halo stars, the cooling phenomenon of halo metal-poor stars can explain their current disk population, and has corollaries for chemical evolution of disk galaxies in general.
\end{abstract}

\keywords{Galaxy dynamics(591) --- Galaxy bars(2364) --- Galaxy disks(589) --- Galaxy evolution(594) --- Galaxy formation(595)}


\section{Introduction}
\label{sec:intro}

Recent observations have detected a population of low metallicity stars with [Fe/H]$\le $-2.5\,dex in the Milky Way (MW) stellar disk. 31\% of these stars reside in the disk, $|z|\le 3$\,kpc of its midplane \citep[e.g.,][]{sestito20,lucey21}. {\it Pristine} photometric survey \citep{stark17}, Hamburg/ESO \citep{reimers90,beers99} and Gaia EDR3 \citep{gaia20} surveys quantified some of the properties of this population. It appears to be statistically biased to populate prograde (with disk rotation) orbits. \citet{bella24} have analyzed metal-poor stars in the Solar neighborhood and found that among the disky kinematics, the ratio of the prograde to retrograde orbits is $\sim 3/1$. \citet{zhang23} have analyzed the Gaia DR3 kinematic sample of the metal-poor population and found two kinematic groups --- one stationary and one with a net prograde rotation of $80\,\mathrm{km\,s^{-1}}$. 

Suggestions have been made to the origin of these metal-poor stars in the disk. They can come from early buildup of the MW, especially retrograde stars, and from minor mergers \citep[e.g.,][]{sestito21,carollo23}. Metal-poor disk stars also appear on very eccentric orbits. Other suggestions include the early formation of these stars entirely in the MW disk that is heated to a thick disk \citep[e.g.,][]{dimatteo20}. Or they could be shepherded by the stellar bar, i.e., trapped by its corotation resonance from the inner galaxy \citep[e.g.,][]{dillamore23,yuan23}. 

Here, we analyze a new process which affects the halo population, moving  its small fraction into the stellar disk, and propose a novel scenario for the origin of the observed low metallicity stars confined close to the mid-plane of disk galaxies. Majority of disk galaxies host stellar bars on various sizes \citep[e.g.,][]{sellwood93,knapen00,jogee04}. These systems have been modeled extensively using 3D numerical simulations, and many of their properties have been successfully analyzed \citep[e.g.,][]{sellwood80,atha83,pfenniger84,weinberg85,combes90,raha91,atha03,marti06,dubinski09,villa09,bi22}. Nevertheless, these collisionless systems continue to provide new effects, e.g., effect of spinning halos and bulges \citep[e.g.,][]{long14,collier18,li23b,li23c}. We define the DM halo spin as $\lambda = J/J_\mathrm{max}$ following \citet{bullock01}, where $J$ is the halo angular momentum and $J_\mathrm{max}$ is maximally allowed $J$.

The 3-D dynamics of barred galaxies was found to be associated with a substantially chaotic phase space, whose fraction correlates with the bar strength \citep{atha83,pfenniger84} and the host halo departure from axial symmetry \citep{elzant02}. In conservative dynamical systems, the size of chaotic region is a measure of the resonance width --- both the concepts of a chaos and a resonance are related. Resonances can overlap, so it becomes impossible to relate a given chaotic orbit to single resonance. In more than 2-D non-integrable potentials, the chaotic space at a given energy is connected via the Arnold's web ``surrounding'' the regular islands of quasi-periodic orbits \citep{licht92}. In a secular evolution, the regular islands may grow or shrink, allowing orbits to migrate in a stochastic way between islands possessing different orbital properties.  

One can partition the phase space in barred galaxies in three distinct regions \citep[see Fig.\,1 in][]{olle98}: first, inside the corotation, at the Jacobi energy lower than the one of the relevant Lagrange points; second, outside the corotation; and third, the rest of the phase space at higher Jacobi energy. The last region involves most of the DM particles, whose motion is allowed throughout the phase space. The fraction of chaotic phase space in the bar neighborhood is expected to be large, as very few regular islands exist, e.g., the retrograde inclined ``anomalous'' family \citep{heisler82}. In particular, the radial orbits along the bar rotation axis with amplitudes exceeding the disk scaleheight are very unstable \citep{martinet87,pfenniger87}, meaning that the phase space close to the rotation axis in the halo is connected to the chaotic web. 

The most chaotic halo orbits are those able to oscillate between large radii and the bar region. Away from the bar, they behave as orbits in the axisymmetric or spherical potentials. When penetrating the bar region they can be scattered  quasi-randomly, subject to bar orientation. Resonances have been found to dominate interactions between the vertical and disk orbits, and are associated with energy transfer away from the disk motions \citep[][]{friedli90,combes90,li23a}. Moreover, interactions between the halo population and the embedded barred stellar disks involve many resonances between halo orbits and stellar bars, connecting widely different phase space regions by the chaotic web. 

In this work, we highlight one of such effects that has eluded the attention of modelers and theorists --- the resonant cooling of vertical stellar oscillations in spinless and $\lambda = 0.06$ halos. The higher value of $\lambda$ has been selected based on the measurement of the MW DM halo spin, $\lambda\sim 0.061-0.088$ \citep{obreja22}.  We limit ourselves to the by-product of this effect --- the origin of disk metal-poor stars. We analyze our results and discuss their implications for galaxy evolution.

%%%%%%%%%%%%%%%%%%%%%%%%%%%%%%
\section{Numerical Methods}
\label{sec:numerics}
%%%%%%%%%%%%%%%%%%%%%%%%%%%%%%

We use the $N$-body part of the mesh-free hydrodynamics code \textsc{gizmo} \citep{hopk15}, an extension of the \textsc{gadget-2} code \citep{sprin05}. The models of stellar disk galaxies embedded in spherical DM halos have been constructed with two different halo spins using methods described elsewhere \citep[e.g.,][]{collier18,li23b}. We re-simulated the model with the halo virial radius of 180\,kpc from \citet{li23b} with $\lambda=0$ and $\lambda=0.06$, increasing the frequency of dumping snapshots to every 1\,Myr, using $N_\mathrm{DM} = 7.2 \times 10^6$ DM particles and $N_\mathrm{S} = 8 \times 10^5$ stellar particles. The mass-per-particle in the disk is $7.9 \times 10^4 \, \mathrm{M_{\odot}}$ and $8.7 \times 10^4 \, \mathrm{M_{\odot}}$ in the halo. The gravitational softening is $\epsilon = 25 \, \mathrm{pc}$ for each particle. The models have been run for $10\,\mathrm{Gyr}$, with total angular momentum conserved within 0.2\% and total energy within 0.1\% over the total integration time. 

The DM component has the NFW density profile \citep[]{nfw96} relaxed in the exponential disk potential, with $r_{\mathrm{c}} = 1.4\,\mathrm{kpc}$, the flat density core, and $r_{\mathrm{s}} = 10\, \mathrm{kpc}$, the characteristic radius. We use a Gaussian cut-off radius $r_{\mathrm{t}} = 180$\,kpc to obtain a finite mass of $M_{\mathrm{h}}=6.3 \times 10^{11} \, \mathrm{M_{\odot}}$. The initial exponential stellar disk has the mass $M_\mathrm{d}=6.3 \times 10^{10} \, \mathrm{M_{\odot}}$ and radial scale-length $R_0 = 2.85$\,kpc, and the scale-height $z_0 = 0.6\,\mathrm{kpc}$. Velocities of disk particles are assigned using the epicycle approximation and asymmetric drift correction \citep{li23b}. Two models of DM halos are abbreviated as P00 (for $\lambda=0$) and P60 (for $\lambda=0.06$). The angular momentum axis of the DM halo in the P60 model agrees with that of the disk.

\begin{figure}
    \center 
	\includegraphics[width=0.5\textwidth]{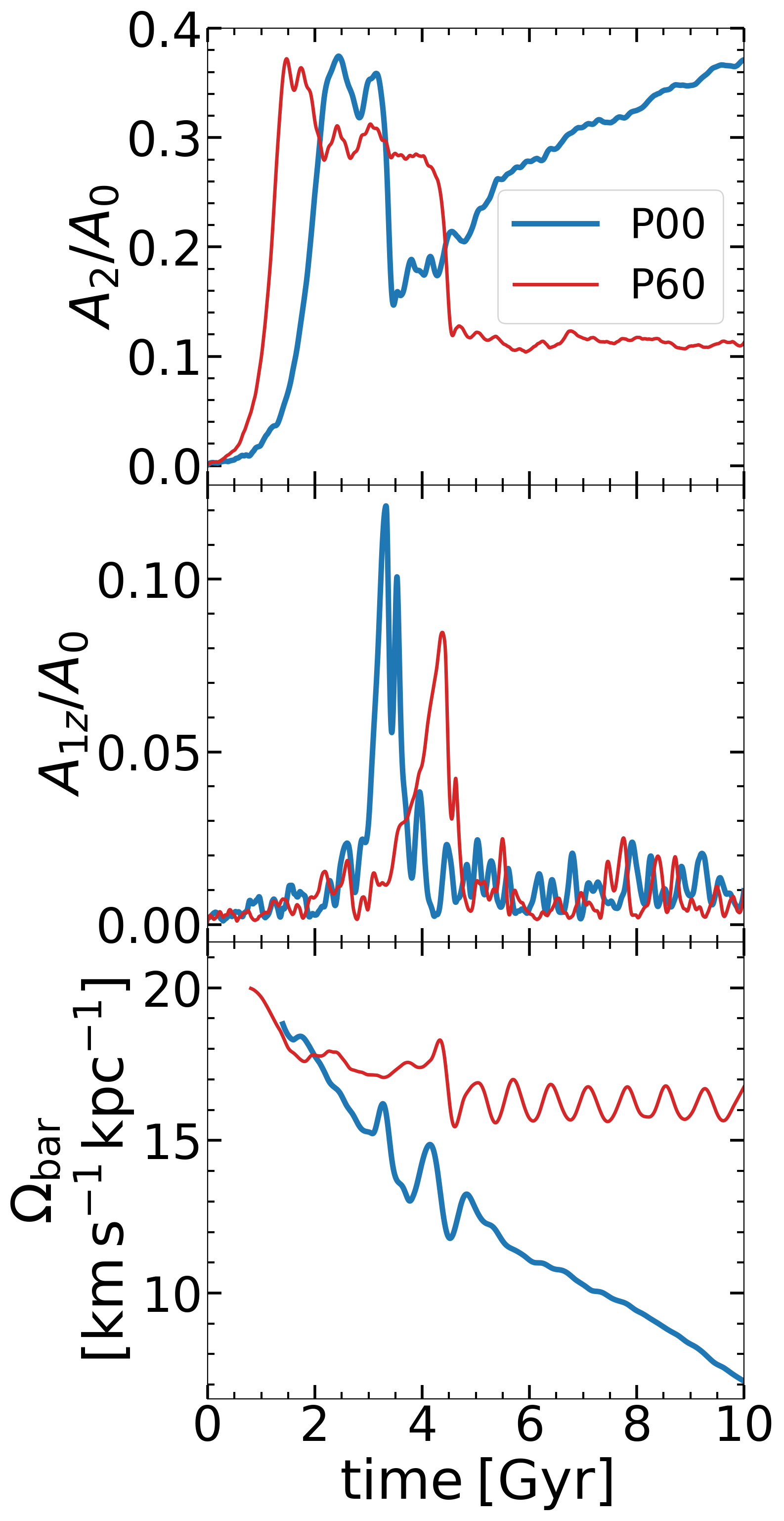}
    \caption{Evolution of the bar strength $A_2$, buckling strength $A_{1z}$, and the bar pattern speed $\Omega_{\mathrm{bar}}$ in the P00 (blue) and P60 (red) models.}
    \label{fig:bar00}
\end{figure}

\begin{figure}
    \center 
    \includegraphics[width=0.6\textwidth]{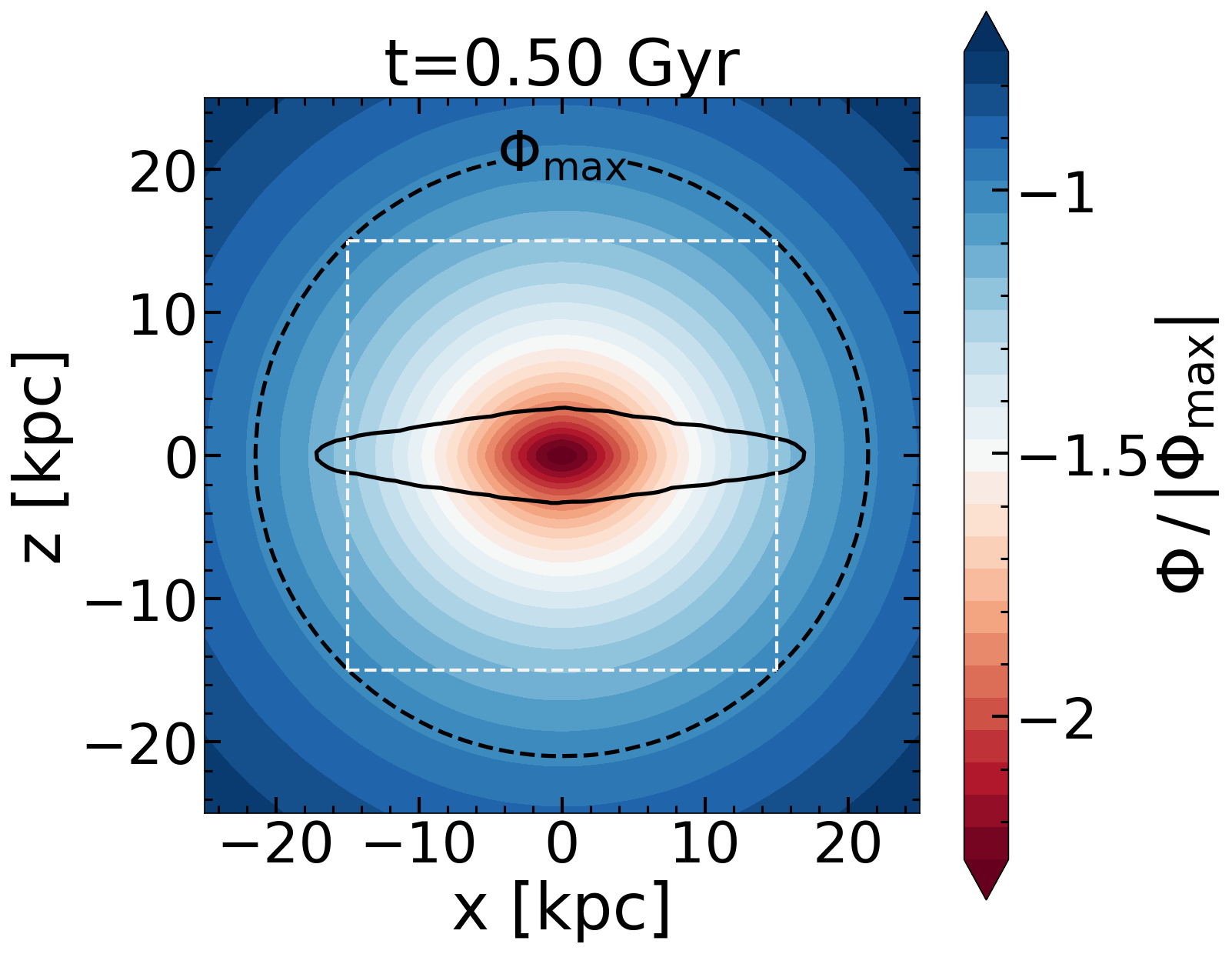}
    \caption{Schematic figure of the region where the DM particles are sampled. The figure has the disk edge-on view ($x$-$z$ plane) at $t=0.5$\,Gyr. The colored contours are the equipotential surfaces of the total system. The gray dashed box indicates the cylinder with a radius 15\,kpc and vertical extension from $z=-15$ to $z=15$\,kpc. The dashed `circle' is an equipotential surface with the maximal potential $\Phi_\mathrm{max}$ found inside the cylinder. The envelope of the disk is plotted as the solid black curve. We randomly sampled $6.5\times 10^5$ DM particles inside $\Phi_\mathrm{max}$ equipotential surface whose specific energy is smaller than $\Phi_\mathrm{max}$.}
    \label{fig:equipot}
\end{figure}

%%%%%%%%%%%%%%%%%%%%%%%%%%%%%
\section{Results}
\label{sec:results}
%%%%%%%%%%%%%%%%%%%%%%%%%%%%%

Both models, P00 and P60, are subject to the bar instability and their evolution is described in \citet{li23b} and in Figure\,\ref{fig:bar00}, displaying the bar Fourier amplitude, $A_2$, the buckling Fourier amplitude, $A_\mathrm{ 1z}$, and the bar pattern speed, $\Omega_\mathrm{ bar}$. The stellar bar evolution depends on the halo $\lambda$. The P00 bar experiences vertical buckling at $t\sim 3.3$\,Gyr, and fully re-grows after, while the P60 bar buckles at about 4\,Gyr,  essentially leaving behind only an oval distortion, and its torque against the DM, therefore, is very weak.  

We have performed an orbit analysis \citep[see e.g.,][for details]{li23c} for both models by constructing a sample of halo particles from each model. The sampled particles have been selected as follows (see Figure\,\ref{fig:equipot}). (1) We calculated the maximal potential $\Phi_{\mathrm{max}}$ from the region within the cylindrical radius $R = 15$\,kpc and vertical distance $|z| < 15$\,kpc at $t = 0.5$\,Gyr; (2) inside the equipotential surface of $\Phi_{\mathrm{max}}$, we randomly choose $6.5\times 10^5$ DM particles whose specific energy is smaller than $\Phi_{\mathrm{max}}$. 

We have determined orbital parameters of DM particles, e.g., their maxima along the $z$-axis, $z_\mathrm{ max}$, and along the cylindrical $R$-axis, $R_\mathrm{ max}$. By focusing on the evolution of $z_\mathrm{ max}(t)$, we have detected its decrease by a factor $\ge 2$ over various time. To analyze statistical properties of these orbits, we have binned evolution of $z_\mathrm{ max}(t)$ into $\Delta t = 0.5$\,Gyr. Next, we have calculated $\zmax$ --- the time-averaged $z_{\mathrm{max}}$ for each consecutive 0.5\,Gyr bin from $t=0$ to $t=10$\,Gyr.  We call the abruptly `cooled' particles those whose $\zmax$ decreases by a factor of $\ge 2$ over $\Delta t\le 0.5$\,Gyr, to separate them from all other cooled halo particles over longer time periods, i.e.,  0.5--1.0\,Gyr, 1.0--1.5\,Gyr, and so on. Note that the $z$-coordinate of each particle has been measured with respect to the Laplace surface defined by the surface where the vertical acceleration $a_{z}$ vanishes \citep[e.g.,][]{dekel83,li23c}.  

\begin{figure}
    \center
    \includegraphics[width=0.45\textwidth]{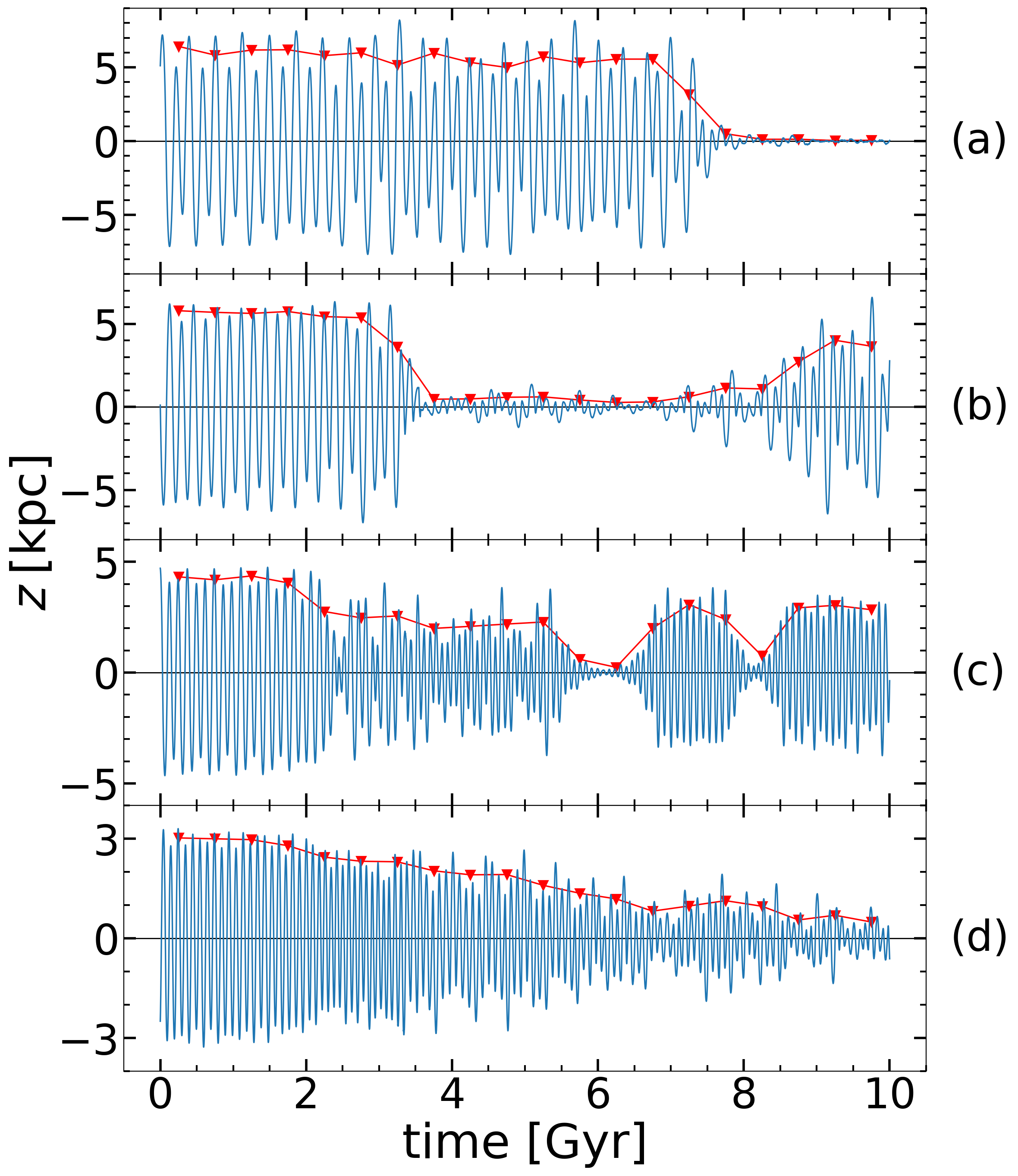}
    \caption{Four representative examples of cooling particles in the P00 model. The $z$-coordinate of each particle is measured with respect to the Laplace surface. $\zmax$ calculated in each 0.5\,Gyr bin has been plotted using red curves. (a) Abruptly cooling particle at $\sim 7.5$\,Gyr and never heats up again until the end of the run. (b) The particle cools down abruptly at $\sim 3.5$\,Gyr and heats up at $\sim 8$\,Gyr. (c) The particle experiences multiple coolings at $\sim 2.5$, $\sim 5.8$, and $\sim 8$\,Gyr. (d) The particle cools down gradually from $\sim 2$ to $\sim 10$\,Gyr.}
    \label{fig:examples_P00}
\end{figure}

\begin{figure}
    \center
    \includegraphics[width=0.4\textwidth]{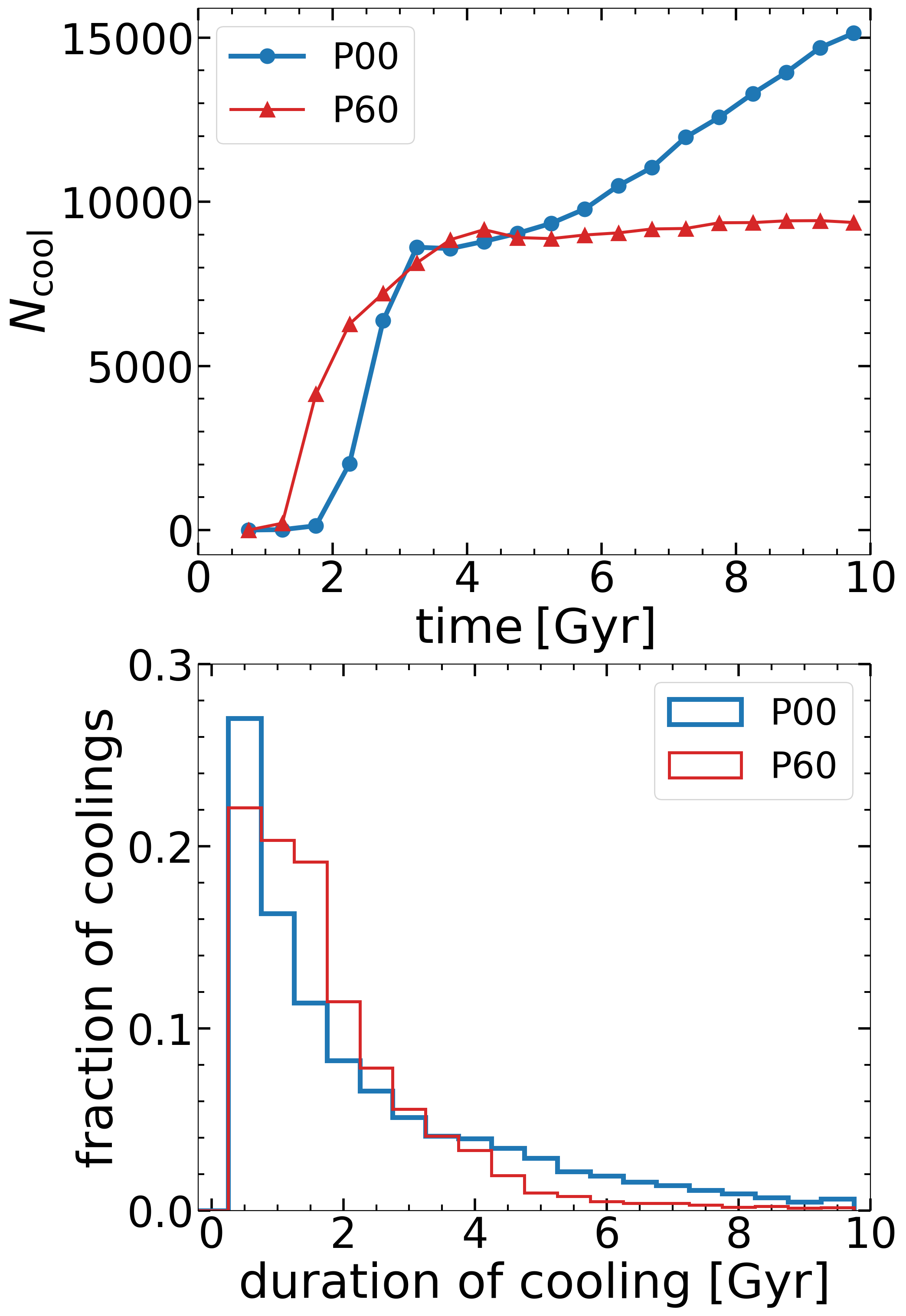}
    \caption{Statistics of a sub-sample of DM halo particles which cool down from high altitude into the disk plane. These particles are selected under two conditions (see text for details) in the P00 (blue) and P60 (red) models. {Top frame:} The \textit{net} number of sub-sampled particles whose $\zmax$ are smaller than the disk thickness after a cooling event. Particles that are heated up and leaving the disk in each time bin are excluded. {\it Bottom frame:} The histogram of the duration of individual cooling events for particles that are found in the disk at $t=10$\,Gyr (i.e., the particles at the end-point in the top panel).}
    \label{fig:cool_to_disk}
\end{figure} 
  
Figure\,\ref{fig:examples_P00} displays the typical behaviors of $\zmax$ of cooling DM particles. Firstly, a particle cools down abruptly by a factor of $\ge 2$ and never heats up again. Second, a particle cools down and heats up after some time period. Next, a particle cools down abruptly and heats up multiple times over the simulation run. Lastly, a particle cools down gradually over 8\,Gyr.

To obtain the statistics of cooled halo particles, we proceed with the following definitions. First, we assume that the stellar disk extends to $z_\mathrm{ disk}=3$\,kpc above/below the midplane, based on the edge-on disk density contours. Second, we focus on these halo particles which acquire the disk kinematics.  Third, in Section\,\ref{sec:discuss}, we consider a \textit{fraction} of cooled DM particles as the low metallicity halo stars that have obtained the disk kinematics. For this purpose, we limit our statistical analysis to DM particles with $\zmax \ge 2z_\mathrm{ disk}$ only before the cooling process begins. This also explains the requirement of cooling $\zmax$ by a factor of $\ge 2$.

We define a sub-sample of all DM halo particles that cool down from the high altitude into the disk in each bin by all types of cooling, i.e., abruptly, gradually, and multiple coolings, $N_\mathrm{ cool}(t)$. In both the P00 and P60 models, the particles of this sub-sample are filtered out of the total sample discussed above by imposing the two conditions: (1) $\zmax > 2 z_\mathrm{ disk}$ at the time of sampling ($t = 0.5 \: \mathrm{Gyr}$); (2) $\zmax < z_\mathrm{ disk}$, after the cooling. 

The top panel of Figure\,\ref{fig:cool_to_disk} shows the cumulative number of cooled DM halo particles as a function of time, $N_\mathrm{cool}(t)$, i.e., only the DM particles that remain in the disk. The re-heated DM particles, i.e., those leaving the disk, have been excluded. Hence, $N_{\mathrm{cool}}(t=10\,\mathrm{Gyr})$ gives the number of the cooled DM particles remaining in the disk by the end of the run. The bottom panel shows the duration distribution of individual cooling events from particles that are found in the disk at $t = 10$\,Gyr.

The sample of DM particles used to analyze the statistics of cooling events includes only the particles with $\zmax$ larger than two times $z_\mathrm{ disk}$ at the time of sampling ($t=0.5$\,Gyr). This number is $N_\mathrm{ total} = 325,270$ for the P00 model and 325,019 for the P60 model. To calculate this number, we have determined the total number of sampled DM particles inside the equipotential surface $\Phi_\mathrm{ max}$. Then corrected it by the fraction of {\it bound} particles with $\zmax\ge 2z_\mathrm{ disk}$.

%%%%%%%%%%%%%%%%%%%%%%%
\section{Discussion}
\label{sec:discuss}
%%%%%%%%%%%%%%%%%%%%%%%

We have shown that population of halo particles that oscillate about the Laplace surface, can decrease their amplitudes of oscillations in a short time period, $\sim 2-3$ oscillations. In other words, they are `cooled down' abruptly. Other DM halo particles cool down but over longer time periods. When associating the DM particle trajectories with those of the halo metal-poor stars, one deals with the migration of halo stars to the disk. A number of interesting corollaries emerge as a result of this new process. But first of all, what is the underlying mechanism of this migration? 

We start with an analogy: a population of ping-pong balls above a moving floor. Both the balls and the floor can oscillate only along the vertical coordinate $z$, each with a different frequency and amplitude. We assume that the ball-floor collisions are elastic and that the ball population is released at various heights above $z=0$. Now, we analyze the results of the ball-floor collisions. 

\begin{table}
\centering
\caption{Statistics of cooled halo particles for P00 and P60 models at $t=10$\,Gyr and of cooled halo stars at redshift $z=0$. $N_\mathrm{ total}$ is the number of sampled DM particles with $\zmax > 2z_{\mathrm{disk}}$ from the entire sample at $t=0.5$\,Gyr; $f_{\mathrm{cool}}$(abrupt) is the probability of finding an \textit{abruptly} cooled halo `star' in the disk, following the disk kinematics; $f_{\mathrm{cool}}$(total) is the probability of finding a cooled halo `star' in the disk due to all types of coolings; P/R is the prograde-to-retrograde ratio of cooled DM particles found in the disk.  $N_{*,\mathrm{cool}}$(total) is the estimated number of halo stars cooled to the disk (see text), using the MW stellar halo.}

\label{tab:tab2}
\begin{tabular}{|l|l|r|r|} 
\hline
                             & Model                         & P00             & P60              \\ 
\hhline{|====|}
\multirow{4}{*}{Simulations} & halo spin $\lambda$           & 0               & 0.06             \\
                             & $N_\mathrm{total}$            & 325,270         & 325,019          \\
                             & $f_\mathrm{cool}$(abrupt)     & 2.9\%           & 1.8\%            \\
                             & $f_\mathrm{cool}$(total)      & 4.7\%           & 2.9\%            \\ 
\hline
Stellar halo                 & $N_{*, \mathrm{cool}}$(total) & $1.2 \times 10^7$ & $7.6 \times 10^6$  \\
\hline
\end{tabular}
\end{table}

Balls will acquire/lose kinetic energy from the collision with the floor, depending on their velocity differences with the floor at the bouncing time. Increasing/decreasing vertical kinetic energy is analogous to `heating up' or `cooling' the balls. Some collisions will result in a substantial cooling of the vertical motions --- this sub-population of balls will be trapped close to the floor. Some of the balls will be trapped for a prolonged time period, some will be re-heated and re-cooled multiple times. These outcomes demonstrate the possibility of the resonance cooling of the balls. 

This toy model of ping pong balls -- moving floor interactions can serve as a simplified analogy to vertical oscillations of halo stars and the horizontal motion of the stellar bar in the embedded galactic disk and their energy exchange. Halo stars can exchange energy and angular momentum with the spinning bar depending on their relative position before and after crossing the bar. As a result, the amplitude of the vertical oscillations will increase or decrease. A resonance between halo orbits and the bar perturbations leads to a fast energy transfer, i.e., during 2--3 vertical oscillations and cooling of these motions, especially because in 3-D potential the chaotic web related to resonances connects different phase space regions. This is the essence of the mechanism operating in barred galaxies and leading to the formation of the `cold' population of metal-poor stars in the disk.  

\begin{figure}
   \centering
    \includegraphics[width=0.7\textwidth]{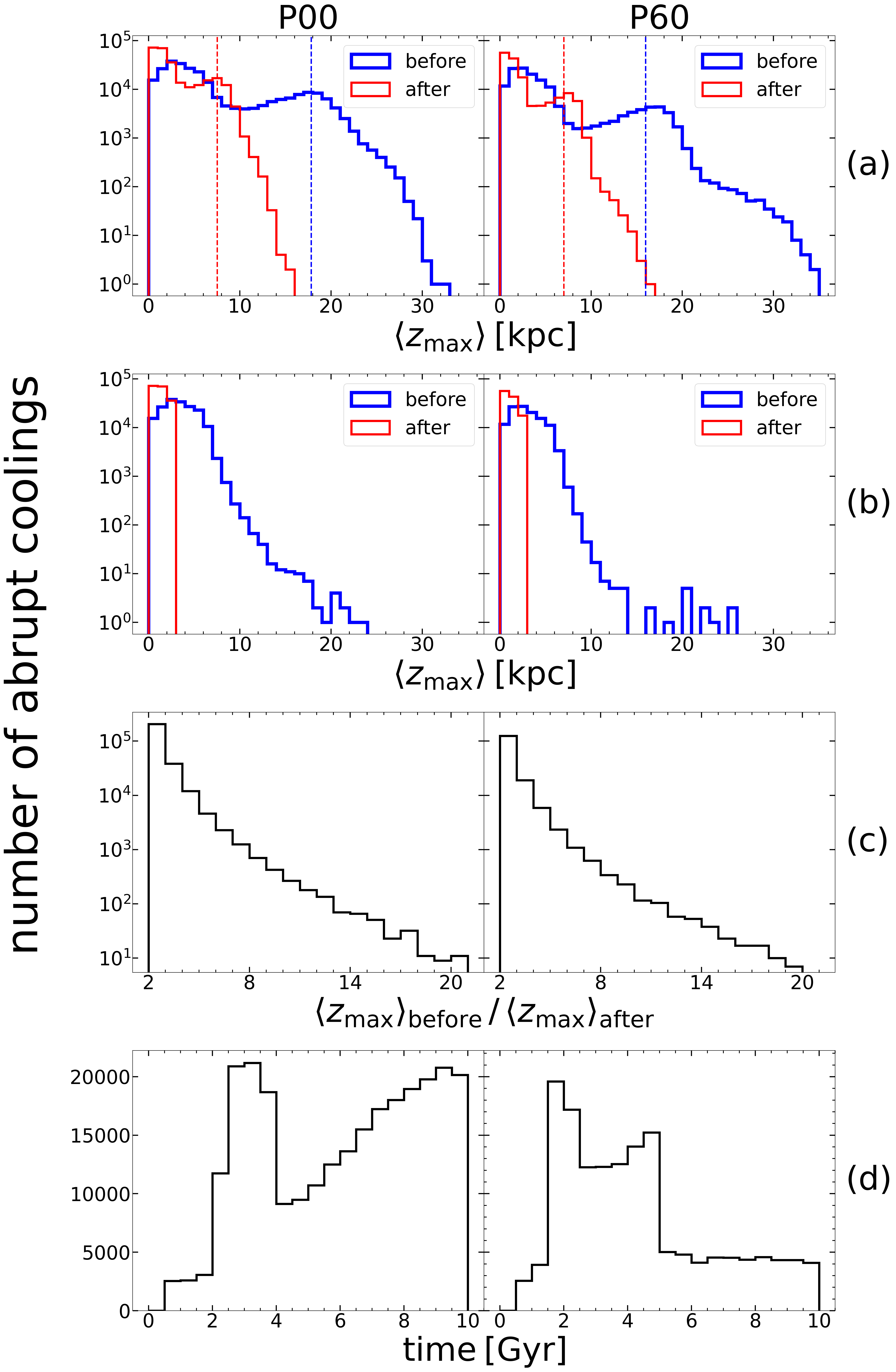}
    \caption{Additional statistics of $\zmax$ for DM halo particles before and after each {\it abrupt} cooling, i.e., in $\le 0.5$\,Gyr, with the P00 (left) and P60 (right). Here, by definition, $\zmax_{\mathrm{before}} \: / \: \zmax _ {\mathrm{after}} > 2$. (a) Distribution of the number of abrupt coolings. The blue (orange) vertical dash line indicates the 90th percentile of $\zmax_{\mathrm{before}}$ ($\zmax_{\mathrm{after}}$) of abrupt coolings. The 90\% of coolings lie to the left of the vertical lines. (b) Distribution of abrupt coolings with $\zmax_{\mathrm{after}} < 3 \: \mathrm{kpc}$. (c) Distribution of $\zmax_{\mathrm{before}} \: / \: \zmax _ {\mathrm{after}}$ of all abrupt coolings. (d) Evolution of the number of abrupt coolings in each time bin.}
    \label{fig:abr_cool_zmax_P00}
\end{figure}

In the P00 model, $\sim 2.67 \times 10^5$ abrupt coolings happened for $\sim  1.84 \times 10^5$ DM particles. In the P60 model, $\sim 1.54 \times 10^5$ abrupt coolings happened for $\sim  1.05 \times 10^5$ DM particles. Figure\,\ref{fig:abr_cool_zmax_P00} displays additional statistics of the {\it abrupt} coolings only. The top frame compares the initial distribution of  $\zmax_{\mathrm{before}}$ with that after the cooling event, $\zmax _ {\mathrm{after}}$.  The next frame shows the distribution of $\zmax _ {\mathrm{after}} < 3$\,kpc for the abrupt coolings. The next frame shows the distribution of the ratios  $\zmax_{\mathrm{before}} / \zmax _ {\mathrm{after}}$ for this type of cooling. Finally, the bottom frame exhibits the evolution of the number of abrupt coolings binned in 0.5\,Gyr bins. 

The empirical probability of a halo particle to cool down to the disk plane and to remain within the disk at the end of the run is 
\begin{equation}
    f_{\mathrm{cool}} = \frac{N_{\mathrm{cool}}}{N_{\mathrm{total}}}
    \label{eq:fcool}
\end{equation}
where $N_\mathrm{ cool}$ given by Figure\,\ref{fig:cool_to_disk}. Results for both models are given in Table\,\ref{tab:tab2}. $f_{\mathrm{cool}} \sim 4.7\%$ in P00, and $f_{\mathrm{cool}} \sim 2.9\%$ in P60.

The above results summarize the statistics of DM halo particles that experience the `cooling' effect over various cooling intervals. To single out the extreme cooling effect, which proceeds over only 0.5\,Gyr, we display the probability of such an {\it abrupt} event, $f_{\mathrm{cool}}$(abrupt), in Table\,\ref{tab:tab2}. An example of such an abrupt cooling has been demonstrated in Figure\,\ref{fig:examples_P00}a.

Observing the evolution of $N_\mathrm{ cool}(t)$ in the top frame of Figure\,\ref{fig:cool_to_disk}, a trend emerges that it is correlated with the amplitude evolution of the stellar bar. In both P00 and P60 models, this number increases until the buckling, i.e., $t \sim 3$\,Gyr in P00, and $t \sim 4$\,Gyr in P60. After the buckling, the bar regrows in the P00 model until the end of the simulation, as is  $N_\mathrm{ cool}(t)$. However, the remaining oval distortion in the P60 model after buckling (Figure \ref{fig:bar00}) leads to a flattened  $N_\mathrm{ cool}(t)$ in P60.

\begin{figure}
    \centering
    \includegraphics[width=0.7\textwidth]{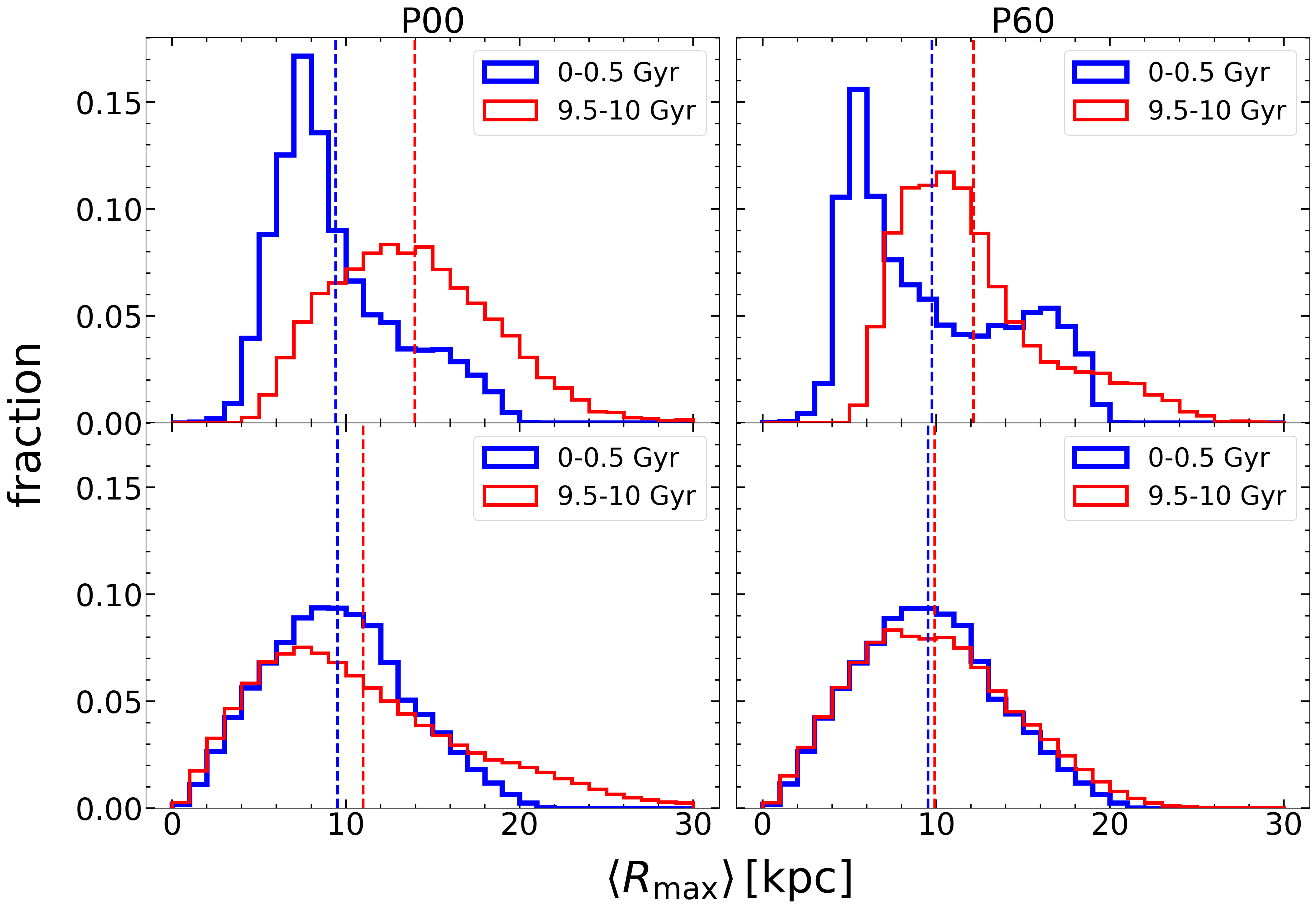}
    \caption{{\it Top:} The distribution function of $\langle R_{\mathrm{max}} \rangle$ from cooled particles of $N_{\mathrm{cool}}(t=10\, \mathrm{Gyr})$ (see Figure \ref{fig:cool_to_disk}). {\it Bottom:} Same as above, but for the entire sample. The vertical dashed lines show the average values for each distribution.}
    \label{fig:Rmax}
\end{figure}

So is the cooling mechanism a resonant one? We easily find that for some DM particles the abrupt cooling correlates with the azimuthal angle between the particle and the stellar bar --- when the particle crosses the Laplace plane slightly ahead of the bar having the same azimuth. This points to a resonant effect, when the energy of the vertical oscillation is transferred to the stellar bar. More generally, we rely on the structure of the phase space, as discussed in the section\,\ref{sec:intro}. The abrupt cooling is impossible if one thinks in terms of the regular phase space, ignoring the properties  of  the chaotic web in 3-D non-integrable potentials. The chaotic web does present many options of these transitions.  Our search for cooling particles is a selective one, because we are interested in this specific aspect of evolution. Disk particles heating up into the halo could also be investigated, as well as other transitions between the distinct dynamical particle groups.

We now turn to estimate the cooling effect on the vertical {\it stellar} oscillation and supplement this with their radial migration for the halo stars in a full analogy with the DM halo particles. For this purpose, we assume that the halo stars have identical mass and are all metal-poor. Using, for example, the \citet{chabrier03} initial mass function and the observed masses of metal-poor halo stars of $0.8-1\,M_\odot$ \citep[e.g.,][]{valentini19}, we take the average stellar mass as $M_\odot\sim 1\,M_\odot$ for simplicity. Note also that the statistics of cooled stars will be affected by replacing the DM particles used in the simulations by this average stellar mass. 

In order to replace the DM particles by the stellar ones, we use the observations of the Milky Way stellar halo. \citet{deason19} estimated stellar mass of the MW stellar halo at $\sim 1.4 \times 10^9 \, \mathrm{M_{\odot}}$. The density of stellar halo follows the power law of $\rho(r) \propto r^{-2.4}$ within $\sim 17.2$\,kpc, and $\rho(r) \propto r^{-4.5}$ in the range $17.2 < r < 97.7$\,kpc \citep[e.g.,][]{kafle24}. 

Hence, the total stellar mass within the equipotential surface $\Phi_{\mathrm{max}}$ in our model, and outside the disk with $|z| > 6$\,kpc, including only the bound stars, can be estimated as $M_\mathrm{*,total}\sim 2.6\times 10^8\,\mathrm{M_{\odot}}$ --- this was calculated exactly as $N_\mathrm{total}$ for the sampled DM particles, explained in section\,\ref{sec:results}. Consequently, using the same cooling probability $f_\mathrm{cool}$ in eq.\,(\ref{eq:fcool}) for the halo stars as for the DM, we obtain $N_\mathrm{*,cool}\sim 1.2\times 10^7$ in the P00 model, and $7.6\times 10^6$ in the P60 model (Table\,\ref{tab:tab2}). Hence, we expect the above $N_\mathrm{*,cool}$ residing in the contemporary MW disk.

\begin{figure}
    \center
    \includegraphics[width=0.5\textwidth]{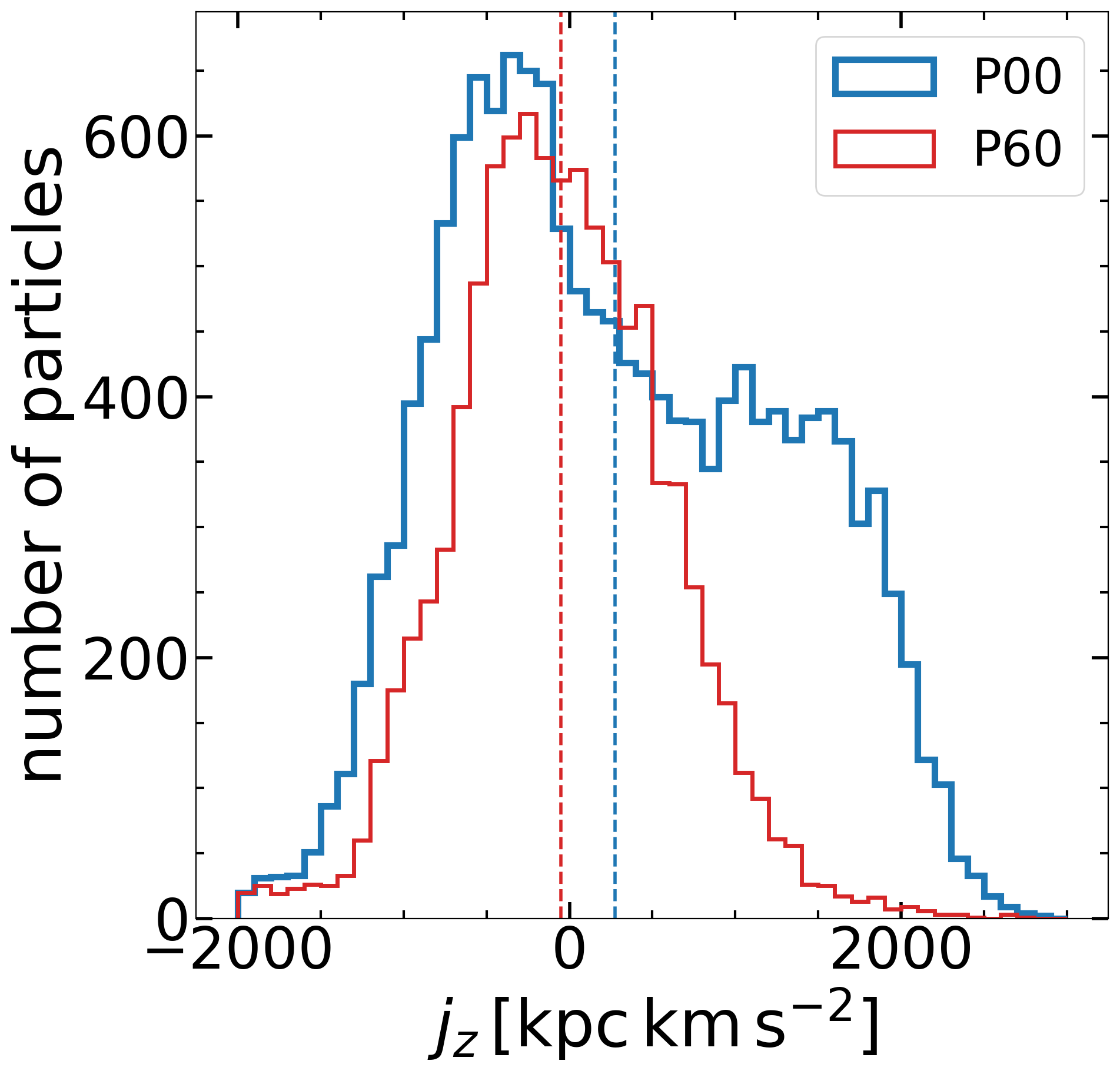}
    \caption{Distribution of the specific angular momentum, $j_\mathrm{ z}$, from cooled particles of $N_{\mathrm{cool}}(t=10\, \mathrm{Gyr})$ (see Figure \ref{fig:cool_to_disk}) for the P00 and P60 models. The vertical dashed lines show the average values for each distribution.}
    \label{fig:jz_t10}
\end{figure} 

Next, we analyze various kinematic parameters of the cooled stellar halo population, such as their radial migration and the ratio of prograde-to-retrograde orbits in the disk. We have measured  $R(t)$ of these particles and calculated $\Rmax$ the same way as $\zmax$. The resulting mean value of the distribution of their $\Rmax$ is 9.4\,kpc at $t=0$\,Gyr and 13.0\,kpc at $t=10$\,Gyr, in the P00 model. For P60 model, the mean value of $\Rmax$ is 9.7\,kpc at $t=0$\,Gyr and 12.1\,kpc at $t=10$\,Gyr. In comparison, the mean value of $\Rmax$ of the entire sampled DM particles at $t=0$\,Gyr and $t=10$\,Gyr shifted from 9.5 to 11.0\,kpc in the P00 model and from 9.5 to 9.9\,kpc in P60 model (Fig.\,\ref {fig:Rmax}).

To obtain the ratio of prograde-to-retrograde orbits of the cooled stellar halo population, we assume that the halo star kinematics follows that of the DM, which is questionable for the P60 when applied to the MW ---  the inner stellar MW halo discussed here has been found to be essentially nonrotating \citep[e.g.,][]{freeman12,das16,deason17}. Under these conditions,  we calculate the instantaneous angular momentum, $J_z$, of cooled particles. The prograde orbits are those spinning with the disk.  The resulting prograde-to-retrograde ratio, P/R, of these particles is $1.2:1$ in the P00 model and $0.8:1$ in the P60 model. However, these average ratios have little significance because the ratio depends on the radius, as we discuss below.

\begin{figure}
    \centering
    \includegraphics[width=0.8\textwidth]{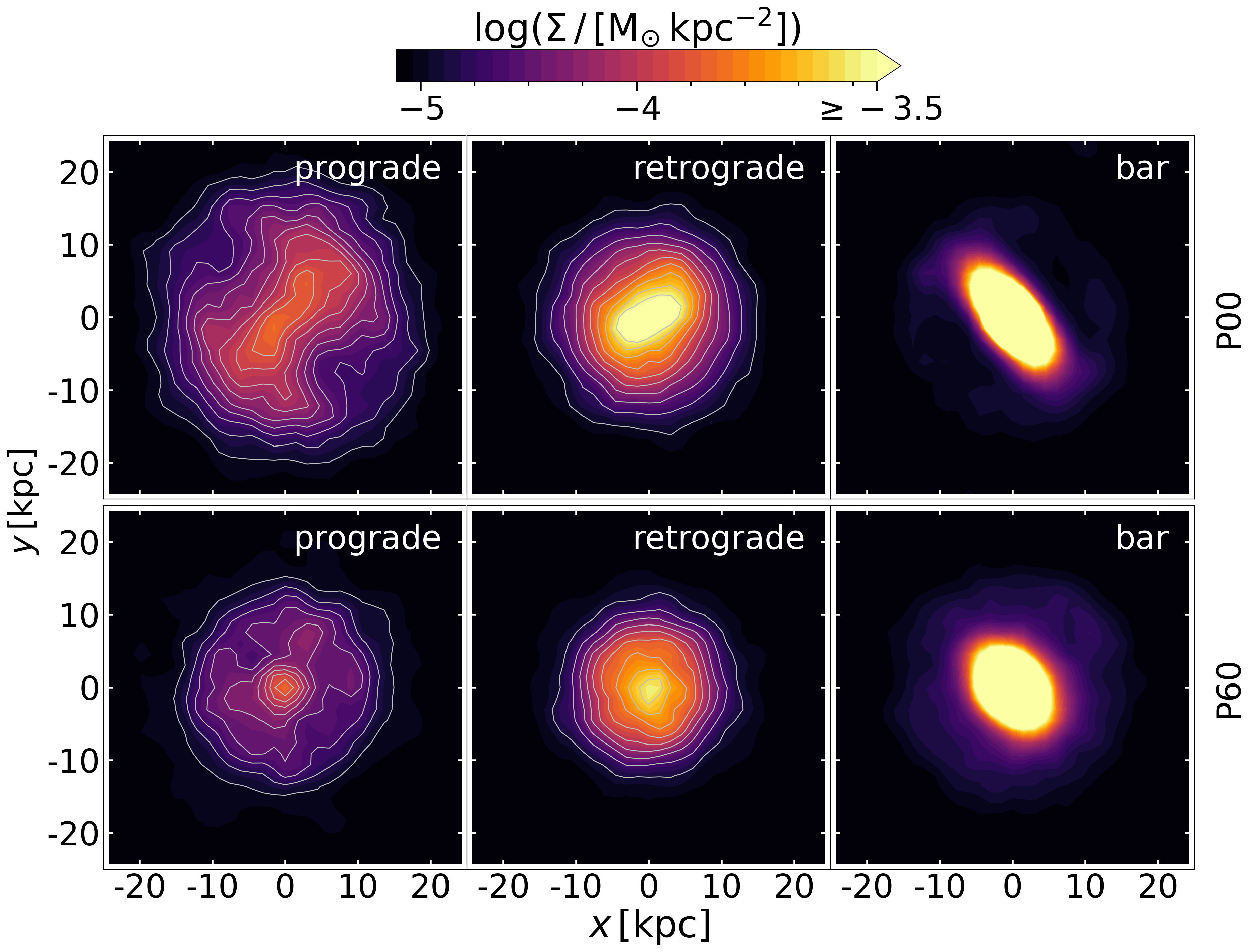}
    \caption{The distribution of cooled DM particles of $N_{\mathrm{cool}}$($t=10$\,Gyr) (see Figure \ref{fig:cool_to_disk}) in the $x$-$y$ plane separated into prograde (left column) and retrograde (middle column) particles, for the P00 (top) and the P60 (bottom) models. The right column is part of randomly sampled disk particles showing the orientation of the stellar bar at the same time. The surface density of particles is plotted using contours with a linearly scaled color palette. The prograde and retrograde orbits have been identified by their instantaneous angular momentum $J_\mathrm{ z}$.}
    \label{fig:pro_retro_contour}
\end{figure}

First, we find that the prograde and retrograde particles are distributed differently in the $xy$-plane at the end of the run (Fig.\,\ref{fig:pro_retro_contour}).  In the P00 model, most of the prograde particles are distributed in a dumbbell shape with a major axis perpendicular to the major axis of the stellar bar.  This dumbbell distribution reaches the radius beyond the bar-size radius. In contrast, most retrograde particles dwell within the bar-size radius and are distributed in an axisymmetric fashion. In the P60 model, with only an oval distortion at the end, the prograde particles have not shown any preferred azimuthal orientation. They are distributed more axisymmetrically than the prograde particles in the P00 model and extend to the region outside the bar-size radius as well. The retrograde particles are distributed similarly to those in the P00 model. Most of them are within the bar-size radius and distributed in an axisymmetric fashion.

\begin{figure}
    \centering    \includegraphics[width=0.7\textwidth]{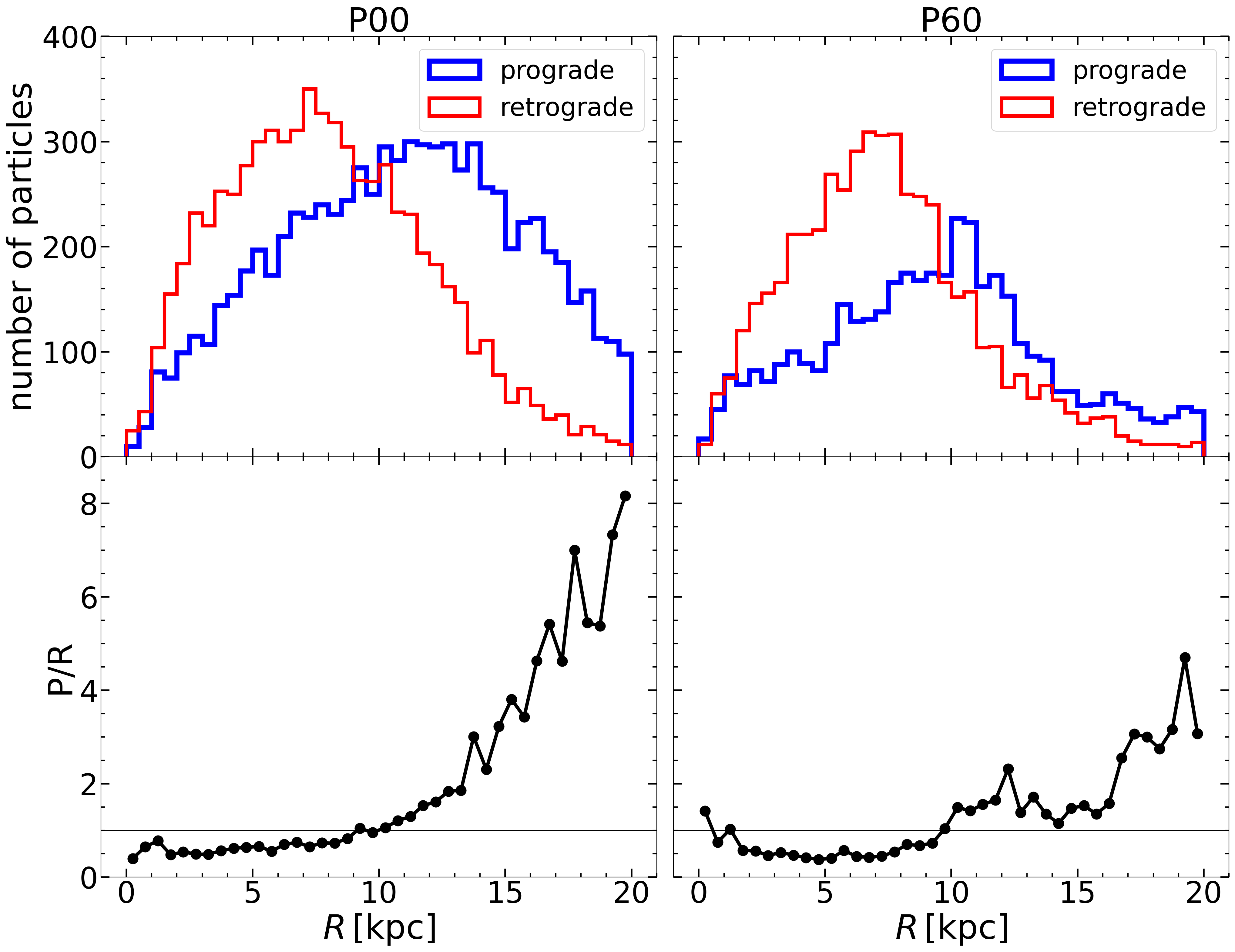}
    \caption{{\it Top:} Radial migration of cooled particle distribution for P00 (left) and P60 (right) models at $t=10$\,Gyr (the same population as in Figure \ref{fig:pro_retro_contour} ). {\it Bottom:} Prograde-to-retrograde orbits ratio, P/R, as a function of radius for the P00 model (left frame) and P60 model (right frame) at $t=10$\,Gyr.}
     \label{fig:ProRetro}
\end{figure}

Second, we checked the radial dependence of the P/R.  Outside the bar radius, in the P00 model, this ratio increases monotonically and becomes larger than 10 outside the corotation radius.  In the P60 model, this ratio is not monotonic but reaches $\sim 2-2.5$ outside the corotation, then drops to unity, following a substantial increase at larger radii. Therefore, faster spinning {\it stellar} halos produce a more shallow distribution of P/R orbits outside the corotation.

No data exists for the radial distribution of the P/R for metal-poor stellar population in the disk. Observations indicating that P/R being $\sim 3:1$ for Gaia DR3 \citep{bella24} and other samples \citep[e.g.,][]{reimers90,beers99,stark17,sestito20,lucey21} may agree with our models. Detailed observations of the MW can provide the radial distribution of metal-poor stars, and observations of other disk galaxies can reveal the scatter in this distribution which probably depends on the assembling history of their stellar halos. 

In summary, we have presented a novel mechanism of cooling the vertical oscillations of halo stars in barred galaxies with nonspinning and spinning DM halos. Using numerical simulations, we argue that this mechanism is associated with a chaotic transition between regular regions separated by the chaotic phase space generated by the strong bar perturbation. And the chaotic web does allow a large choice of possible transitions between the regular islands of phase space.  Finally we remark that the particle cooling presented here is not contradicting the conservative nature of the $N$-body systems, because globally the kinetic energy lost by the particle subsample is transferred to the rest of the system.  This energy transfer,  more generally the action, would be much slower if the system were close to integrable, like spherical systems.    

This mechanism is also associated with the radial migration of metal-poor stars, which acquired disk kinematics and has corollaries to the chemical evolution of disk galaxies. This cooling mechanism acts on a range of timescales, from a dynamical time of $\ltorder 0.5$\,Gyr associated with 2--3 vertical oscillations to the secular timescale of a few Gyr, as expected for a chaotic transition. Using a typical MW stellar halo inferred from recent observations, we estimate the stellar metal-poor population in the contemporary MW disk contributed by this new process. Furthermore, analyzing the kinematics of this disk population, we estimate the radial distribution of the prograde-to-retrograde orbits ratio for the metal-poor disk population and find that the faster spinning stellar halo produces shallower distribution outside the corotation radius. More detailed numerical model comparison with the MW stellar halo properties can provide observational constrains on the stellar halo assembly and the bar evolution.

%% Also note that the akcnowlodgment environment does not support long amounts of text. If you have a lot of people and institutions to acknowledge, do not use this command. Instead, create a new 
% \section{Acknowledgments}.

\begin{acknowledgments}
We thank Phil Hopkins for providing us with the latest version of GIZMO, and to Da Bi for providing the data for stellar halo from his cosmological simulations. I.S. is grateful for a generous support from the International Joint Research Promotion Program at Osaka University.  Simulations have been performed using the University of Kentucky Lipscomb Computing Cluster. We thank Vikram Gazula at the Center for Computational Studies at the University of Kentucky for help with the technical issues during the LCC runs.
\end{acknowledgments}

\bibliography{paper}{}
\bibliographystyle{aasjournal}



\end{document}